\documentstyle[tighten,prb,epsf,floats,aps]{revtex}

\begin{document}

\twocolumn[
\hsize\textwidth\columnwidth\hsize\csname @twocolumnfalse\endcsname

\draft \preprint{\today}
\title{Incoherent Interplane Conductivity of
{$\kappa$-(BEDT-TTF)$_2$Cu[N(CN)$_2$]Br}}

\author{J. J. McGuire, T. R\~o\~om\cite{Room}, A. Pronin\cite{Pronin},
T. Timusk}
\address{Department of Physics and
Astronomy, McMaster University, Hamilton, Ontario L8S 4M1, Canada}

\author{J. A. Schlueter, M. E. Kelly, A. M. Kini}
\address{Chemistry and Materials Science Divisions, Argonne National
Laboratory, Argonne, Illinois 60439, USA}

\maketitle

\begin{abstract}
The interplane optical spectrum of the organic superconductor
{$\kappa$-(BEDT-TTF)$_2$Cu[N(CN)$_2$]Br} was investigated in the
frequency range from 40 to 40,000~cm$^{-1}$.  The optical
conductivity was obtained by Kramers-Kronig analysis of the
reflectance.  The absence of a Drude peak at low frequency is
consistent with incoherent conductivity but in apparent
contradiction to the metallic temperature dependence of the DC
resistivity. We set an upper limit to the interplane transfer
integral of $t_b^2/t_{ac}\approx10^{-7}$~eV. A model of
defect-assisted interplane transport can account for this
discrepancy.  We also assign the phonon lines in the conductivity
to the asymmetric modes of the ET molecule.
\end{abstract}

]

Of the $\kappa$-type (BEDT-TTF)-based organic superconductors
$\kappa$-(BEDT-TTF)$_2$Cu[N(CN)$_2$]Br has the highest
superconducting transition temperature at ambient
pressure,\cite{Kini90} $T_c=12.5$~K. BEDT-TTF, or
bis(ethylenedithio)tetrathiafulvalene and hereafter further
abbreviated to ET, is a platelike molecule that, in the
$\kappa$-type ET-based superconductors, forms conducting layers of
orthogonally arranged dimers separated by insulating layers of
anions.\cite{Kini90}  The resulting quasi-two-dimensional
electronic structure makes this class of organic superconductors
of great interest due to their potential similarity to the high
temperature cuprate superconductors.\cite{McKenzie97} One
important issue is the nature of interlayer transport: is it
coherent giving rise to a three dimensional Fermi liquid at
sufficiently low temperature, or does it remain incoherent to the
lowest temperatures? This issue is well illustrated by two
two-dimensional superconductors, Sr$_2$RuO$_4$ which shows
coherent interplane transport at low temperature and
Bi$_2$Sr$_2$CaCu$_2$O$_{8+\delta}$ (Bi-2212) where the interplane
transport is incoherent at all temperatures.

The first, Sr$_2$RuO$_4$, is isostructural with the cuprate
superconductor La$_{2-x}$Sr$_x$CuO$_4$, and has a superconducting
$T_c$ of 1.5~K.\cite{Maeno94} Although highly anisotropic, it
shows coherent interplane transport. At low temperature
Sr$_2$RuO$_4$ shows metallic resistivity both in the in-plane and
interplane directions, although the resistivity in the interplane
direction is two orders of magnitude higher than that of the
in-plane direction and has a metallic temperature dependence only
below 100~K.\cite{Maeno94} The in-plane optical conductivity shows
a Drude-like peak, and a weak Drude component can also be seen in
the interlayer direction.\cite{Katsufuji96,Hildebrand00}

The cuprate superconductor Bi-2212 shows an in-plane DC
resistivity that, in optimally doped Bi-2212, varies linearly with
temperature up to several hundred Kelvin, while in the interplane
direction the resistivity is four orders of magnitude higher and
{\it increases} with decreasing
temperature.\cite{Watanabe96,Martin88} The real part of the
optical conductivity of Bi-2212 shows a Drude-like peak centred
at zero frequency in the in-plane direction,\cite{Quijada99} but no
such peak is seen in the interplane direction.\cite{Tajima93}
Instead, the interplane conductivity consists entirely of phonon
lines.  The weak residual interplane transport is due to an
incoherent tunneling mechanism.\cite{Watanabe96}

The transport properties of $\kappa$-(ET)$_2$Cu[N(CN)$_2$]Br are
also highly anisotropic.\cite{Buravov92,Dressel94} Within the
conducting $ac$-planes, the resistivity shows an unusual broad
peak near 100~K, but has a clearly metallic behaviour at low
temperature with resistivity decreasing with decreasing
temperature. The interplane $b$-axis resistivity has qualitatively
the same metallic temperature dependence as the in-plane
resistivity, but is three orders if magnitude higher.  Its value
of 1~$\Omega$cm at 15~K, which depends somewhat on how quickly the
sample is cooled,\cite{Su98} is six orders of magnitude higher
than what is seen in good metals. Thus, in the interplane
direction, there is an apparent contradiction between the
temperature dependence of the resistivity that suggests coherent
transport, and the magnitude of the resistivity that suggests
incoherent transport.

One criterion for evaluating the coherence of the interplane
conductivity is the Ioffe-Regel-Mott minimum metallic
conductivity. Modified for the open Fermi surfaces of these highly
anisotropic systems,\cite{Xie92} it gives a lower limit on
conductivity which corresponds to a coherence length comparable to
the size of a unit cell in the interlayer direction:
\begin{equation}
\sigma_{min}\approx \sqrt{\sigma_\bot\over{\sigma_\Vert}}{e^2\over{2\pi^2\hbar}}
{l_\bot\over{l_{\Vert 1}l_{\Vert 2}}}
\label{sigmin}
\end{equation}
where $\sigma_\bot$ and $l_\bot$ are the conductivity and lattice
constant in the interplane direction and $\sigma_\Vert$, $l_{\Vert
1}$ and $l_{\Vert 2}$ are the conductivity and lattice constants
in the planes.  Table~\ref{sigtab} lists these parameters for
Sr$_2$RuO$_4$, Bi-2212 and $\kappa$-(ET)$_2$Cu[N(CN)$_2$]Br. The
conductivity anisotropy ratio should be evaluated at
$\sigma_\bot=\sigma_{min}$. For Sr$_2$RuO$_4$ this gives coherent
conductivity below 52~K which is indeed the temperature where the
resistivity starts to deviate from the low-temperature $T^2$
dependence. For Bi-2212 and $\kappa$-(ET)$_2$Cu[N(CN)$_2$]Br,
however, $\sigma_{min}>\sigma_\bot$ even near $T_c$ where the
anisotropy is largest. This suggests that the low temperature
interplane conductivity of Sr$_2$RuO$_4$ is coherent but becomes
incoherent above 52~K due to thermal fluctuations, while both
Bi-2212 and $\kappa$-(ET)$_2$Cu[N(CN)$_2$]Br remain incoherent at
all temperatures.

\begin{table}[t]
\caption{Ioffe-Regel-Mott minimum metallic
conductivity for highly anisotropic systems.
$\l_{\Vert 1}$, $l_{\Vert 2}$ and $l_\bot$ are the two in-plane and one
interplane room temperature lattice constants\cite{Maeno94,Hazen90,Kini90}
in \AA, $T$ is temperature in Kelvin, and $\sigma_\Vert$, $\sigma_\bot$
and $\sigma_{min}$ are the in-plane, interplane and minimum conductivities
in $(\Omega cm)^{-1}$ at temperature $T$.}
\begin{tabular}{r|ccccccc}
& $\l_{\Vert 1}$ & $l_{\Vert 2}$ & $l_\bot$ & $T$ & $\sigma_\Vert$
& $\sigma_\bot$ & $\sigma_{min}$\\
\hline
\\
Sr$_2$RuO$_4$ & 3.9 & 3.9 & 12.7 & 52 & 28600 & 37 & 37\\
Bi$_2$Sr$_2$CaCu$_2$O$_{8+\delta}$ & 3.8 & 3.8 & 30.9 & 85 & 25000 & 0.17 & 3.4\\
$\kappa$-(ET)$_2$Cu[N(CN)$_2$]Br & 12.9 & 8.5 & 30.0 & 12.5 & 1000 & 1 & 11\\
\end{tabular}
\label{sigtab}
\end{table}

Optical conductivity provides another method of investigating the
coherence of interplane transport where in a coherent system the
optical conductivity $\sigma_1$ shows a Drude peak centred at
zero frequency with a maximum corresponding to the DC conductivity
$\sigma_{DC}$ and a width equal to the scattering rate $\Gamma$ of
the free carriers:
\begin{equation}
\sigma_1(\omega)={{\sigma_{DC}\Gamma^2}\over{\omega^2+\Gamma^2}}
\label{drude}
\end{equation}
The question of the coherence of the interplane transport in the
ET-based organic superconductors can be addressed with
measurements of low temperature far-infrared optical conductivity
in the interplane direction. To date however, infrared studies
have focused on the in-plane properties,
\cite{Eldridge91,Sugano89,Kornelsen91,Kornelsen92,Meneghetti86,Kaplunov85,Jacobsen85}
and the few interplane measurements which have been made
\cite{Tokumoto88,Vlasova91,Vlasova92,Vlasova93} were at room
temperature and above 600~cm$^{-1}$.  This is due to the
difficulty of growing crystals with large faces perpendicular to
the conducting planes. Recently, however, high quality crystals of
sufficient size for far-infrared interplane measurements have
become available.  The interplane reflectance measurements of
$\kappa$-(ET)$_2$Cu[N(CN)$_2$]Br presented in this paper are the
first such measurements on any ET based superconductor.

The single crystals of $\kappa$-(ET)$_2$Cu[N(CN)$_2$]Br were synthesized
by the electrocrystallization technique described elsewhere.\cite{Kini90}
Typical crystal sizes were $1.5\times1.5\times1.5$~mm with faces as large as
1~mm$^2$ parallel to the interlayer $b$-axis.  Polarized reflectance
measurements between 40 and 8000~cm$^{-1}$ were performed on these as-grown
faces with a Michelson interferometer using three different detectors.  A
grating spectrometer with three additional detectors was used to make
measurements at 300~K for the rest of the range up to 40,000~cm$^{-1}$ (5~eV).

\begin{figure}[t]
\leavevmode
\epsfxsize=\columnwidth
\centerline{\epsffile{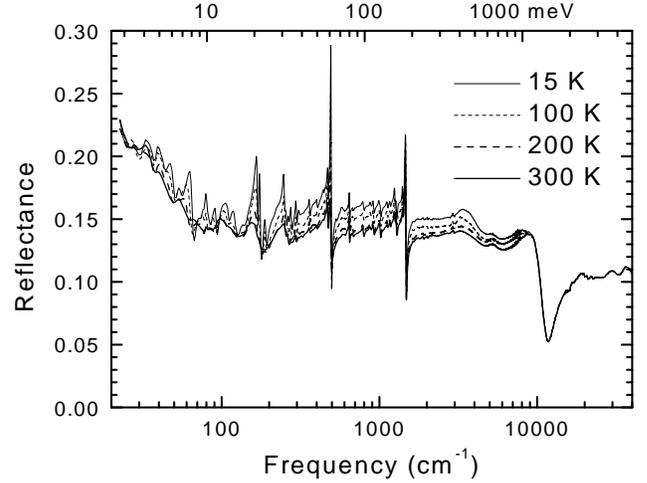}}
\vspace{0.1in}
\caption{Semilog plot of the interlayer reflectance of
$\kappa$-(ET)$_2$Cu[N(CN)$_2$]Br.}
\label{refl}
\end{figure}

The reflectance of $\kappa$-(ET)$_2$Cu[N(CN)$_2$]Br with the light polarized
in the interplane direction is shown in Fig.~\ref{refl} for four
temperatures above the superconducting transition temperature.  The
reflectance is approximately 0.15 over the entire range with
several sharp phonon peaks at low frequencies and some broader interband-like
features at higher frequencies.

\begin{figure}[t]
\leavevmode
\epsfxsize=\columnwidth
\centerline{\epsffile{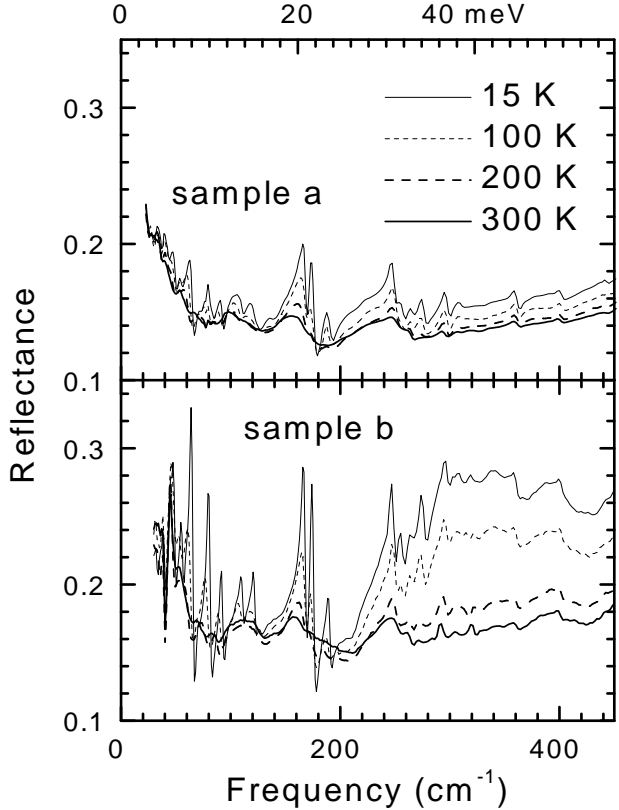}}
\vspace{0.1in}
\caption{A comparison of the interlayer reflectance of two crystals
of $\kappa$-(ET)$_2$Cu[N(CN)$_2$]Br.  The one in the lower panel shows
stronger temperature dependence of the background above 200~cm$^{-1}$.}
\label{variation}
\end{figure}

It should also be pointed out that there was some sample-to-sample
variation of the interplane reflectance, particularly the
temperature dependence of the background reflectance above
200~cm$^{-1}$.  Fig.~\ref{variation} is a comparison of the sample
shown in Fig.~\ref{refl} (upper panel) with another sample showing
a much stronger temperature dependence (lower panel).  The phonon
lines below 200~cm$^{-1}$ also appear to be stronger in this
second sample although the optical conductivity should be
calculated for a true comparison. Unfortunately we were unable to
measure reflectance above 800~cm$^{-1}$ for this second sample,
and we have no explanation for the variation.  Complete data sets
need to be collected on more samples to properly investigate this
phenomenon.  It is interesting to note that the feature has the
appearance of a gap, and evidence for a pseudogap in
$\kappa$-(ET)$_2$Cu[N(CN)$_2$]Br at $T^*=50$~K in
ESR\cite{Kataev92} and $^{13}$C NMR\cite{Mayaffre94,Kawamoto95}
measurements has been reported. This would give
$2\Delta/k_BT^*=5.7$ which is not far from the value 4.3 reported
for the high temperature cuprate superconductors.\cite{Nakano98}

The calculation of the optical conductivity
requires extrapolation of the reflectance to all frequencies for the
Kramers-Kronig analysis.  The spectra were extrapolated to high frequencies
using power law extrapolations: $\omega^{-1}$ above 5~eV and
$\omega^{-4}$ above 62~eV. Below our lowest measured point at 25~cm$^{-1}$
we used a Drude extrapolation, although an insulating extrapolation gives very
similar results. We estimate our experimental
uncertainty of the reflectance to be $\pm0.005$. Combined with uncertainties
due to the extrapolations, this gives an uncertainty in our optical
conductivity of $\pm 8\%$ between 200 and 2000~cm$^{-1}$. Outside this range
the uncertainty rises, reaching $\pm 40\%$ at 25 and 5000~cm$^{-1}$. The
resolution of the spectra is 2~cm$^{-1}$ up to 200~cm$^{-1}$, 4~cm$^{-1}$ up
to 680~cm$^{-1}$, and 15~cm$^{-1}$ up to 8000~cm$^{-1}$.

\begin{figure}[t]
\leavevmode
\epsfxsize=\columnwidth
\centerline{\epsffile{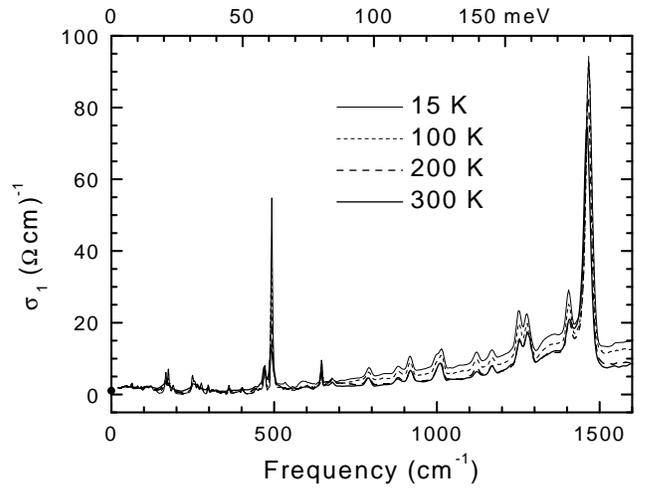}}
\vspace{0.1in}
\caption{The real part of the interlayer optical conductivity of
$\kappa$-(ET)$_2$Cu[N(CN)$_2$]Br. The DC value is marked
on the vertical axis at 1~$(\Omega cm)^{-1}$.}
\label{sig1}
\end{figure}

The real part of the interplane optical conductivity is shown in
Fig.~\ref{sig1}.  The value of the DC conductivity from the work
of Su {\it et al.}\cite{Su98} at 15~K is shown at 1~$(\Omega
cm)^{-1}$ on the vertical axis. Clearly there is no sign of the
usual Drude peak that accompanies coherent conductivity. Instead
the conductivity is dominated by sharp phonon lines on a
background due to interband-like features in the mid-infrared.
This confirms that the interlayer conductivity in
$\kappa$-(ET)$_2$Cu[N(CN)$_2$]Br is incoherent.

A fit to Lorentz oscillators which will be discussed later shows
no sign of any free carrier component to the optical conductivity
above 30~cm$^{-1}$. This is similar to what has been reported for
Bi-2212.\cite{Tajima93}  It is difficult to estimate an upper
limit for the plasma frequency of a hypothetical Drude peak, but
assuming the width is equal to the in-plane width reported by
Eldridge and Kornelsen of $\Gamma=20$~cm$^{-1}$ we estimate
$\omega_p=50$~cm$^{-1}$ for the plasma frequency of the Drude
component. Combining the in-plane scattering rate
$\Gamma=20$~cm$^{-1}$ with the DC resistivity gives a somewhat
lower value of $\omega_p=25$~cm$^{-1}$. Using the 50~cm$^{-1}$
value and a simplified tight-binding model we can estimate the
interplane transfer integral using\cite{Jacobsen85}
\begin{equation}
\omega_p^2={e^2\over\epsilon_0\hbar^2}\sum_{BZ}f(E_{\bf k})
{\partial^2E\over\partial k^2_\mu}
\label{wp}
\end{equation}
where $f(E_{\bf k})$ is the Fermi-Dirac occupation number and the derivative
is to be taken in the direction of the field.  We use the simplified
tight-binding band
\begin{equation}
E_{\bf k}=-2t_a\cos(k_ad_a)-2t_b\cos(k_bd_b)-2t_c\cos(k_cd_c)
\label{band}
\end{equation}
where $d$ are the ET molecular repeat distances and $t$ are the average
transfer integrals along the various directions.  We assume an open Fermi
surface in the interplane direction\cite{Jacobsen83} with $t_b\ll t_{ac}$ and
ignore the in-plane anisotropy to get
\begin{equation}
\omega_{pb}\approx{e^2\over\epsilon_0\hbar^2}
{b^2/V_m\over\sqrt{\pi}\sin(\sqrt\pi)}{t_b^2\over t_{ac}}
\label{wpb}
\end{equation}
where $b$ is the interplane lattice constant and $V_m$ the volume per ET
molecule. This gives $t_b^2/t_{ac}\approx10^{-7}$~eV which
gives
$t_b\approx10^{-4}$~eV if $t_{ac}\approx10^{-1}$~eV as has
been estimated
for (ET)$_2$I$_3$.\cite{Jacobsen85}  Of course, this analysis assumes a
coherent component to the conductivity which seems unlikely given the earlier
discussion of minimum metallic conductivity.

We now return to the problem of the temperature dependence of the interplane
DC conductivity which shows a clear metallic character and in fact follows
rather accurately the in-plane conductivity although it is a factor of 1000
smaller. Since the interplane conductivity is below the minimum metallic
limit it is highly likely that its metallic temperature dependence is due to
some special process that makes it mirror the in-plane conductivity. There
are several models that can do this\cite{Kumar92,Forro92} and we focus here
on a model originally proposed for Bi-2212 by Martin {\it et
al.},\cite{Martin88} where the conducting planes are connected by a
random network of shorts approximated as a regular array of links
of resistance $R_{lb}$ distance $\xi$ apart along the conducting
plane. Given the resistance $R_{lac}$ along the plane between the shorts
and the interplane distance $d$, the apparent interplane
resistivity is given by
\begin{equation}
\rho_b=(R_{lb}+R_{lac})\times{(\xi/2)^2\over d}
\label{rhob}
\end{equation}
and assuming $R_{lac}=\rho_{ac}/d$ we get for the anisotropy
\begin{equation}
{\rho_b\over\rho_{ac}}=\biggl(1+{dR_{lb}\over\rho_{ac}}\biggr)\times
\biggl({\xi\over2d}\biggr)^2
\label{rhob/ac}
\end{equation}

\begin{figure}[t]
\leavevmode
\epsfxsize=\columnwidth
\centerline{\epsffile{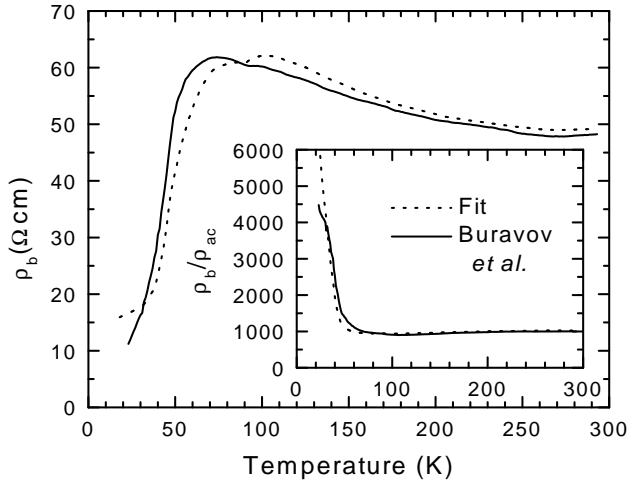}}
\vspace{0.1in}
\caption{A comparison of the resistivity data of Buravov {\it et al.}
\cite{Buravov92} to a fit to the model of Martin {\it et al}.
\cite{Martin88}}
\label{linkfit}
\end{figure}

Buravov's data\cite{Buravov92} above 50~K are nearly temperature-independent
with $\rho_b/\rho_{ac}\approx 1000$, while below 50~K the
anisotropy rises dramatically.  Fig.~\ref{linkfit} compares the interplane
resistivity and anisotropy of Buravov {\it et al.} with a least squares
fit using Eq.~\ref{rhob/ac} assuming a temperature-independent $R_{lb}$.
The fit, which reproduces both the temperature independence of the
anisotropy at high temperatures and the dramatic increase in anisotropy
at low temperatures, gives $R_{lb}=138\pm9$~k$\Omega$ and
$\xi=802\pm7$~\AA~if we take $d=15$~\AA.  The model explains the rise of the
anisotropy below 50~K as due to the rapid drop of $R_{lac}$ relative to the
temperature-independent $R_{lb}$, whereas models where the proportionality
between in-plane and interplane resistivity is built in\cite{Kumar92,Forro92}
predict a constant anisotropy, independent of temperature. The observed
sample-to-sample variation shown in Fig.~\ref{variation} is also consistent
with a process controlled by defects.

The model of a meandering current will fail at high frequency. To
estimate the characteristic maximum frequency we model the current
path as a transmission line of series resistors $R_{lac}$ shunted
to ground by capacitors $C_s$ where we set $C_s \approx 2\pi\xi$.
The impedance of such a transmission line is
\begin{equation}
Z={R_{lac}\over2}\Biggl(1+\sqrt{1+{4i\over\omega R_{lac}C_s}}\Biggr)
\label{Z}
\end{equation}
Thus above a frequency
given by $\omega_0 = (R_{lac}C_s)^{-1}$ the capacitors short the AC
current and the in-plane fields do not have time to build up. With
the parameters determined above we find that $\omega_0 \approx 10$~cm$^{-1}$.
A test of the model would be a reduction of the interplane conductivity from
its DC value to a much lower value at this frequency. Dressel {\it et al.}
\cite{Dressel94} find in the millimeter wave range (1-3~cm$^{-1}$)
conductivities that agree with DC values consistent with our picture.
Unfortunately the strong phonon background discussed in the next paragraphs
prevents us from giving an accurate value of the interplane electronic
conductivity in the far infrared range other than an upper limit
of about 1~($\Omega$cm)$^{-1}$.

It was possible to fit the conductivity with a series of Lorentzian
oscillators according to
\begin{equation}
\sigma_1(\omega)={1\over{4\pi}}\sum_i{{\omega_{pi}^2\omega^2\Gamma_i}\over
{(\omega_{0i}^2-\omega^2)^2+\omega^2\Gamma_i^2}}
\label{sigma}
\end{equation}
51 oscillators were used in the fit along with a weak Drude term to account
for phonons below the measurement range and one strong oscillator at high
frequency to provide the background tail of the mid-infrared features.  It
was necessary to add some asymmetry to the strongest oscillators at 492 and
1450 cm~$^{-1}$, but all other oscillators were symmetric.  Table~\ref{osc} lists
the frequency $\omega_0$, plasma frequency $\omega_p$ and width $\Gamma$ of
the oscillators at all four temperatures. Most of the low frequency lines
seem to be lattice modes as their frequencies increase with decreasing
temperature.\cite{Dressel92}  Some of these lines were previously reported
in a powder absorption experiment\cite{Dressel92} and are marked with an
asterisk. The rest of the lines are related to internal
vibrations of the ET molecule, and these mode assignments were made by
comparison with Eldridge {\it et al.} who assign the normal modes of the ET
molecule\cite{Eldridge95} and relate these ET modes to observed lines in
in-plane infrared and Raman spectra of $\kappa$-(ET)$_2$Cu[N(CN)$_2$]Br.
\cite{Eldridge96,Eldridge96b}

\begin{table*}[p]
\caption{Lorentz oscillator parameters from least squares fits of
Eq.~(\ref{sigma}) to the real part of the interplane optical conductivity
of $\kappa$-(ET)$_2$Cu[N(CN)$_2$]Br at 15, 100, 200 and 300 K. All values
are in cm$^{-1}$.  Vertical lines associate one or more mode assignments
with a set of oscillators. $\ast$ indicate lines seen by Dressel
{\it et al}.\cite{Dressel92}}
\begin{tabular}{rcccp{5pt}cccp{5pt}cccp{5pt}ccc}
&\multicolumn{3}{c}{15 K}&&\multicolumn{3}{c}{100 K}&&
\multicolumn{3}{c}{200 K}&&\multicolumn{3}{c}{300 K}\\
\cline{2-4}\cline{6-8}\cline{10-12}\cline{14-16}
&$\omega _0$ & $\omega _p$ & $\Gamma$ && $\omega _0$ & $\omega _p$ & $\Gamma$ &&
$\omega _0$ & $\omega _p$ & $\Gamma$ && $\omega _0$ & $\omega _p$ & $\Gamma$\\
\hline
                     &    28.4&   4.3&  1.9&&   27.9&   5.7&  2.4&&   29.4&   7.6&  4.0&&   29.2&   5.5&  3.9\\
                     &    34.4&  15.3&  5.8&&   34.5&  15.7&  6.4&&   34.6&  17.1&  7.2&&   34.0&  17.2&  7.2\\
                     &    41.3&  21.8&  7.5&&   41.0&  19.0&  6.8&&   40.9&  15.0&  5.9&&   41.1&  18.1&  9.3\\
$\ast$               &    47.9&  19.2&  5.6&&   47.0&  21.5&  7.7&&   46.1&  21.9&  8.9&&   46.3&  19.8& 11.5\\
$\ast$               &    55.5&  25.4& 10.8&&   54.7&  21.5&  9.6&&   53.9&  18.8&  9.5&&   54.6&  20.2& 10.9\\
$\ast$               &    63.5&  27.3&  5.7&&   61.6&  27.7&  7.6&&   59.6&  23.6&  8.6&&   59.6&  20.0& 12.4\\
                     &    72.0&  12.4&  4.4&&   71.4&  19.1&  8.4&&   69.9&  27.2& 14.4&&   69.5&  28.5& 15.4\\
                     &    75.7&   5.0&  2.1&&   75.8&   8.1&  2.7&&   75.4&  13.0&  6.0&&   75.3&   8.5&  5.1\\
$\ast$               &    81.2&  30.2&  8.9&&   80.2&  23.3&  8.0&&   80.4&   8.5&  3.8&&   80.3&   5.4&  2.5\\
$\ast$               &    91.3&  20.7&  5.1&&   89.5&  23.1&  8.2&&   86.7&  25.1& 12.8&&   84.8&  17.6& 10.5\\
                     &   103.0&  21.8&  9.4&&       &      &     &&       &      &     &&       &      &     \\
$\ast$               &   109.1&  24.7&  7.6&&  106.3&  31.0& 12.3&&  103.7&  18.6&  9.7&&       &      &     \\
$\ast$               &   121.2&  39.8& 13.3&&  118.9&  30.4& 15.8&&  113.4&  35.3& 27.8&&       &      &     \\
                     &   136.5&  14.7&  7.1&&       &      &     &&       &      &     &&       &      &     \\
$\ast$               &   167.1&  39.7&  4.6&&  166.1&  47.2&  9.5&&  165.0&  52.1& 17.3&&       &      &     \\
$\ast$               &   175.3&  42.9&  4.7&&  175.1&  37.7&  7.5&&  175.9&  35.1& 14.9&&       &      &     \\
$\ast$               &   190.0&  27.7&  5.5&&  190.0&  27.2&  7.4&&  189.2&  22.0& 11.8&&       &      &     \\
\multicolumn{1}{r|}{$\ast$}
                     &   248.7&  39.0&  5.3&&  249.7&  23.7&  4.9&&  251.4&  30.6&  8.9&&       &      &     \\
\multicolumn{1}{r|}{$\ast$}
                     &   255.9&  39.2&  8.6&&  253.2&  51.5& 15.2&&       &      &     &&       &      &     \\
\multicolumn{1}{r|}{$\nu_{53}$(B$_{2u}$)}
                     &   265.6&  21.8&  5.6&&  265.7&  17.1&  4.9&&  265.0&  19.2&  6.2&&       &      &     \\
$\nu_{36}$(B$_{1u}$)$\ast$
                     &   276.3&  32.0&  5.8&&  275.8&  29.3&  6.8&&  275.5&  22.0&  9.8&&       &      &     \\
$\ast$               &   297.7&  24.0&  3.1&&  297.6&  21.4&  3.5&&  296.7&  20.0&  4.8&&  295.3&  20.0&  7.7\\
                     &   311.1&  22.1&  8.0&&  311.7&  16.6&  7.0&&  310.2&  10.5&  5.1&&  308.5&  14.0&  8.6\\
                     &   325.5&  33.3& 26.5&&  321.9&  23.3& 13.7&&  324.6&  29.9& 30.0&&  321.5&  24.1& 18.2\\
$\nu_{52}$(B$_{2u}$) &   360.9&  27.6&  5.2&&  361.6&  20.2&  3.8&&  361.6&  22.5&  6.5&&  361.6&  16.4&  5.9\\
$\nu_{35}$(B$_{1u}$) &   402.2&  28.2&  7.6&&  402.6&  21.9&  6.3&&  402.3&  19.6&  6.8&&  401.4&  17.2&  7.2\\
\multicolumn{1}{r|}{$\nu_{34}$(B$_{1u}$)}
                     &   471.3&  57.8&  6.4&&  471.1&  58.6&  7.5&&  470.3&  63.9&  9.6&&  469.4&  65.4& 11.4\\
\multicolumn{1}{r|}{}
                     &   492.1& 125.5&  4.6&&  491.3& 112.7&  5.2&&  490.4& 101.9&  8.0&&  489.8&  94.3& 12.0\\
$\nu_{33}$(B$_{1u}$) &   595.3&  91.8& 54.2&&  598.9&  68.7& 48.0&&  596.9&  63.0& 54.0&&  595.3&  33.2& 25.1\\
$\nu_{51}$(B$_{2u}$) &   645.8&  48.6&  5.2&&  645.8&  48.1&  5.6&&  644.8&  48.1&  6.4&&  644.1&  44.9&  6.9\\
$\nu_{50}$(B$_{2u}$) &   676.5&  68.5& 40.7&&  677.8&  68.8& 33.6&&  676.7&  63.8& 28.5&&  677.5&  37.4& 19.0\\
$\nu_{32}$(B$_{1u}$) &   791.2&  73.8& 23.4&&  790.5&  62.7& 21.5&&  790.0&  61.6& 21.0&&  788.9&  53.4& 17.4\\
$\nu_{15}$(A$_{u}$)  &   837.7& 105.0& 61.2&&  837.4&  85.8& 56.2&&  836.1&  63.4& 49.3&&  839.2&  38.2& 34.6\\
\multicolumn{1}{r|}{$\nu_{49}$(B$_{2u}$)}
                     &   881.9&  92.2& 34.5&&  881.9&  83.4& 31.9&&  881.0&  69.1& 28.3&&  879.4&  63.9& 29.5\\
\multicolumn{1}{r|}{$\nu_{48}$(B$_{2u}$)}
                     &   917.2&  92.1& 22.0&&  917.8&  85.0& 20.4&&  919.1&  75.9& 20.0&&  919.0&  73.8& 20.5\\
\multicolumn{1}{r|}{$\nu_{31}$(B$_{1u}$)}\\
$\nu_{30}$(B$_{1u}$) &   998.7&  93.7& 28.3&&  999.6&  83.4& 27.3&& 1001.7&  78.0& 27.5&& 1002.8&  78.3& 24.2\\
$\nu_{47}$(B$_{2u}$) &  1015.4&  78.6& 17.8&& 1015.0&  76.4& 17.6&& 1013.7&  61.1& 16.0&& 1012.8&  51.3& 13.1\\
$\nu_{14}$(A$_{u}$)  &  1121.8& 103.9& 31.9&& 1122.2&  97.5& 31.8&& 1122.2&  78.1& 27.6&& 1123.2&  83.1& 38.5\\
$\nu_{67}$(B$_{3u}$) &  1167.3& 113.3& 36.3&& 1168.9& 104.9& 36.3&& 1168.6&  78.6& 27.2&& 1168.3&  76.1& 25.5\\
$\nu_{46}$(B$_{2u}$) &  1251.3& 132.3& 21.1&& 1251.7& 119.3& 20.3&& 1252.5& 100.9& 19.8&& 1253.0&  98.0& 21.2\\
$\nu_{29}$(B$_{1u}$) &  1277.3& 138.2& 24.6&& 1278.2& 131.3& 24.1&& 1278.9& 130.1& 24.6&& 1279.4& 131.1& 26.2\\
\multicolumn{1}{r|}{$\nu_{28}$(B$_{1u}$)}
                     &  1404.4& 134.7& 19.2&& 1405.4& 133.9& 21.2&& 1406.1& 129.4& 24.3&& 1406.9& 123.0& 24.6\\
\multicolumn{1}{r|}{$\nu_{45}$(B$_{2u}$)}\\
$\nu_{27}$(B$_{1u}$) &  1467.7& 343.9& 22.9&& 1467.3& 325.0& 19.9&& 1465.5& 316.9& 20.8&& 1462.5& 321.1& 24.0\\
$\nu_{26}$(B$_{1u}$) &  2921.0&  62.7& 17.8&& 2920.4&  60.9& 18.1&& 2922.4&  55.0& 18.7&& 2920.4&  52.0& 20.2\\
                     &  2938.6&  53.7& 24.0&& 2937.4&  49.4& 25.0&& 2938.9&  48.7& 24.4&& 2939.6&  48.6& 26.0\\
$\nu_{44}$(B$_{2u}$) &  2960.5&  52.5& 20.2&& 2960.8&  58.0& 24.9&& 2963.1&  54.5& 23.2&& 2962.8&  51.4& 20.2\\
$\nu_{66}$(B$_{3u}$) &  2982.4&  44.5& 13.6&& 2983.2&  46.1& 15.6&& 2982.0&  38.6& 18.1&& 2981.3&  48.3& 24.9\\

\end{tabular}
\label{osc}
\end{table*}

In general, infrared spectra are sensitive to asymmetric (ungerade) modes
while Raman spectra are sensitive to symmetric (gerade) modes.  Since the ET
molecule consists of two mirrored halves joined by a single C=C bond, the
infrared and Raman spectra of ET contain similar sets of lines.
\cite{Eldridge95}  Each vibration of atoms in one half of the molecule can
be in phase or out of phase with an identical vibration in the other half
producing symmetric/asymmetric pairs of modes.  Since the two halves are
nearly independent, the members of each mode pair have nearly the same energy.
This argument does not apply to modes involving the central C=C bond.

Like the ET spectra, the in-plane infrared and Raman spectra of
$\kappa$-(ET)$_2$Cu[N(CN)$_2$]Br also have similar sets of lines, however
in this case both sets of lines are ascribed to the symmetric modes of
ET.\cite{Eldridge96}  The reason the symmetric ET modes are infrared-active
in $\kappa$-(ET)$_2$Cu[N(CN)$_2$]Br is the arrangement of ET molecules into
dimers so that the members of a dimer can vibrate out of phase transferring
charge back and forth and producing the dipole moment required for infrared
activity.  The Raman line positions in $\kappa$-(ET)$_2$Cu[N(CN)$_2$]Br are
shifted from the ET line positions due to the charge on the ET molecules,
and the infrared line positions are further shifted by their coupling to the
charge transfer between members of a dimer.\cite{Eldridge96} To assist them
in their mode assignments, Eldridge {\it et al.} also cause further shifts
using several isotopic substitutions.

\begin{figure}[t]
\leavevmode
\epsfxsize=\columnwidth
\centerline{\epsffile{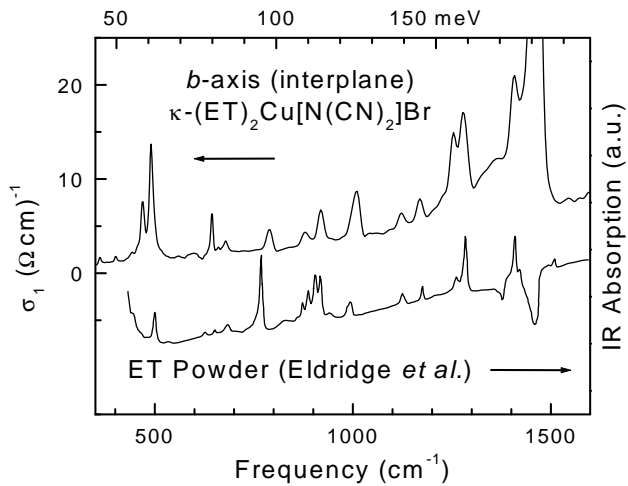}}
\vspace{0.1in}
\caption{A comparison at 300~K of the real part of the interlayer optical
conductivity with the powder absorption spectra of Eldridge {\it et al}.
\cite{Eldridge95}}
\label{powder}
\end{figure}

\begin{figure}[t]
\leavevmode
\epsfxsize=\columnwidth
\centerline{\epsffile{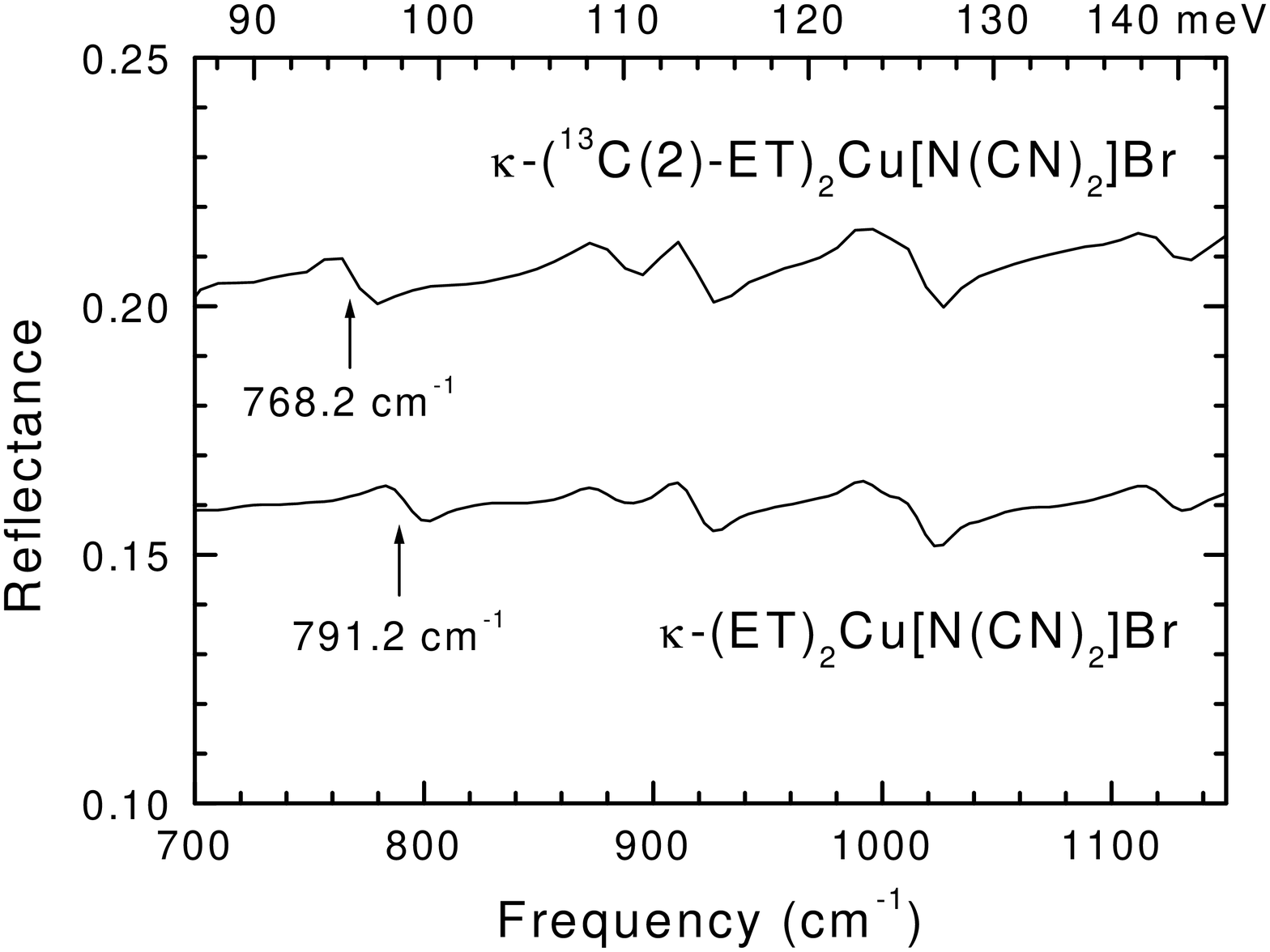}}
\vspace{0.1in}
\caption{A comparison at 15~K of the interlayer reflectance of
$\kappa$-(ET)$_2$Cu[N(CN)$_2$]Br with that of a sample in which the two
central carbon atoms have been substituted with $^{13}$C.}
\label{C2}
\end{figure}

We have assigned the lines in our interplane infrared spectra to
the asymmetric modes of the ET molecule.  The mechanism that makes the
symmetric modes infrared-active in the in-plane measurements does not apply
to our interplane measurements since the electric vector of the light is
perpendicular to the direction of charge transfer between members of a dimer.
Fig.~\ref{powder} shows a comparison of our $\sigma_1$ to the ET powder
absorption spectrum of Eldridge {\it et al}.\cite{Eldridge95}
For the most part, our mode assignments are based on this comparison
since we do not have many measurements on isotopic analogs to assist us.
We do however have a partial reflectance spectrum of a
$\kappa$-($^{13}$C(2)-ET)$_2$Cu[N(CN)$_2$]Br crystal in which the two central
carbon atoms have been substituted with $^{13}$C.  Fig.~\ref{C2} shows the
23~cm$^{-1}$ shift of a line at 791.2~cm$^{-1}$ confirming its identity as
the $\nu_{32}$(B$_{1u}$) mode.\cite{Eldridge95}  Since this mode involves the
central C=C bond, its symmetric counterpart is at a very different frequency
and is seen in the in-plane infrared and Raman spectra near 450~cm$^{-1}$
while no line is seen near 791~cm$^{-1}$.\cite{Eldridge96}

In general, all of the lines become narrower as temperature decreases as
expected, however the plasma frequencies, or line strengths, also increase
significantly in some cases. In particular, the line at 81~cm$^{-1}$, which
is quite strong at 15~K, has nearly disappeared above 200~K.  An increase in line
strength with decreasing temperature in organic conductors has generally
been interpreted as a signature of charge-density-wave fluctuations.
\cite{Rice76,Rice77} However, temperature-dependent phonon intensities have
also been associated with spin-density-wave transitions.\cite{ng84}

In summary we have measured the interplane reflectance of the
quasi-two-dimensional organic superconductor
$\kappa$-(ET)$_2$Cu[N(CN)$_2$]Br and calculated the optical
conductivity using Kramers-Kronig relations. We find strong
evidence of incoherent transport as has been reported for the high
temperature superconductors, and estimate an upper limit for the
free carrier plasma frequency of 50~cm$^{-1}$ from which we derive
an upper limit for the interplane transfer integral of
$10^{-4}$~eV. We have used a defect model to explain the crossover
from a temperature-independent resistivity anisotropy at high
temperatures to a rapidly increasing anisotropy at low
temperatures. We have also fit the phonon lines in the
conductivity to a series of Lorentzian oscillators and assigned
these to the asymmetric modes of the ET molecule.

The work at McMaster University was supported by the
Natural Sciences and Engineering Research Council of Canada.
Work at Argonne National Laboratory is sponsored by the U.S. Department of
Energy, Office of Basic Energy Sciences, Division of Materials Sciences,
under Contract W-31-109-ENG-38.

\end{document}